\documentstyle[12pt,a4,epsf]{article}
\newcommand{\rcite}[1]{{\cite{#1}}}
\newcommand{\rref}[1]{{(\ref{#1})}}
\newcommand{\tref}[1]{{\ref{#1}}}
\newcommand{\rlabel}[1]{{\label{#1}}}
\newcommand{\rbibitem}[1]{\bibitem{#1}}
\newcommand{\be}{\begin{equation}}
\newcommand{\ee}{\end{equation}}
\newcommand{\ba}{\begin{eqnarray}}
\newcommand{\ea}{\end{eqnarray}}
\newcommand{\dis}{\displaystyle}
\begin{document}
\begin{titlepage}
\begin{flushright}
{FTUV/95-70}\\
{IFIC/95-73}\\
{NORDITA-95/84 N,P}\\
{hep-ph/9512374}
\end{flushright}
\vspace{2cm}
\begin{center}
{\large\bf The $\gamma\gamma\to\pi^0\pi^0$ and $\eta\to\pi^0\gamma\gamma$
Transitions in the Extended NJL Model}\\
\vfill
{\bf Johan Bijnens$^a$, Ansar Fayyazuddin$^a$
 and Joaqu\'{\i}m Prades$^b$
}\\[0.5cm]
${}^a$ NORDITA, Blegdamsvej 17,\\
DK-2100 Copenhagen \O, Denmark\\[0.5cm]
$^b$Departament de
 F\'{\i}sica Te\`orica, Universitat de Val\`encia and\\
 IFIC, CSIC - Universitat de Val\`encia,
 C/ del  Dr. Moliner 50, \\ E-46100 Burjassot (Val\`encia),
Spain\\[0.5cm]
\end{center}
\vfill
\begin{abstract}
We calculate within the Extended Nambu--Jona-Lasinio model
the leading in $1/N_c$ contribution to $\gamma\gamma\to\pi^0\pi^0$
to all orders in the external momenta and quark masses.
This result is then combined with the known two-loop
Chiral Perturbation Theory and compared with the data and other calculations.
A technical difficulty in the same calculation beyond order $p^6$ for
$\eta\to\pi^0\gamma\gamma$ is identified and for this decay results
up to order $p^6$ are presented.
\end{abstract}
\vfill
December 1995
\end{titlepage}

Dipion production in photon-photon collisions has been considered a
good test of Chiral Perturbation Theory (CHPT) since its first
calculations\rcite{BC,DHL}. For the neutral  pion case,
its leading contribution is order $p^4$
in the chiral counting and the tree level vanishes at this order.
The size of the prediction for the latter
 case was afterwards confirmed\rcite{CB} but there was
a discrepancy with the predicted behaviour as a function of the center-of-mass
energy. This discrepancy could be understood within the context of
final state scattering effects as was shown using several ways of
unitarizing the lowest order CHPT amplitude\rcite{Pennington,DGH2}.
It has also been calculated within the framework of generalized
CHPT\rcite{Knecht}.

This process ($\gamma \gamma \to \pi^0 \pi^0$)
will be measured with precision equal to or better than
the Crystal Ball data at  DA$\Phi$NE and
other $\Phi$-factories\rcite{DAFNE}.
This prompted the calculation of the next-to-leading correction in CHPT.
This is a two-loop calculation and was performed in
\rcite{BGS}.  The new free parameters appearing in
the tree level contribution (order $p^6$ in this case)
  need to be determined from other processes,
this is at present impossible and leaves an uncertainty in the prediction
{}from this calculation. Another possibility is to estimate them,
here of course model dependence enters and we are leaving pure CHPT.
In the original calculation\rcite{BGS}, this
was done using  resonance exchange dominance
in the same way as was done in
\rcite{ABBC} for the  similar process $\eta\to\pi^0\gamma\gamma$,
see \rcite{longlist,BDV}.
One of the problems appearing in this type of estimates
is that the size and signs of several of
the needed  couplings of resonances  are not well
determined,
still leaving an undesirable uncertainty to the prediction.

In the  process $\eta\to\pi^0\gamma\gamma$, the loop contributions up to
order $p^6$ are suppressed by G-parity or the large kaon mass\rcite{ABBC}.
This was in fact confirmed in an explicit calculation of
the one-loop\rcite{ABBC} and part of the two-loop amplitude to order
$p^6$\rcite{Jetter}.
The latter reference also contains a rather exhaustive
discussion of the present theoretical status of this decay.
 Therefore, the tree level contribution to the amplitude
is  in fact the leading one
and the above uncertainty becomes a dominant part of the uncertainty
on the final result for this process.  We will not treat this process
 in the same detail as  $\gamma \gamma \to \pi^0 \pi^0$
for the reasons given below.

The prediction of higher order coefficients in CHPT from various models
has some history. However, the simplest models are resonance exchange
dominance, the constituent quark-loop model
and the Extended Nambu--Jona-Lasinio (ENJL) model.
It was found in \rcite{BBR} that the ENJL model\rcite{ENJLrevs,Physrep} does
give a good representation of the order $p^4$
coefficients, i.e. the so-called $L_i$ coefficients\rcite{GL}.
In particular
it improved on the description for the parameters in the explicit chiral
symmetry breaking sector (i.e., $L_5$ and $L_8$).
For the corresponding predictions of resonance exchange dominance and
 the constituent quark-loop model see \rcite{EGPR} and \rcite{ERT} and
references therein, respectively.
The constituent  quark-loop prediction for $\gamma \gamma \to \pi^0 \pi^0$
was given in \rcite{BDV}
 and in \rcite{ABBC} for $\eta \to \pi^0 \gamma \gamma$. The
calculation within the ENJL model
of these processes to order $p^6$ was also performed in two recent
papers\rcite{BB,Belkov}. There is some disagreement between them.
In this paper we take the attitude that if the leading contributions
start at rather high order in the chiral expansion even higher orders
might also contribute significantly. This is definitely the
case for $\eta\to\pi^0\gamma\gamma$\rcite{ABBC,Jetter} where restriction
to order $p^6$ or making all order estimates significantly changes the results.
In this Letter  we therefore
use the techniques of \rcite{BP1} to calculate the process $\gamma\gamma
\to\pi^0\pi^0$ to all orders in the chiral expansion to leading order in
$1/N_c$ in the ENJL model. This is equivalent to calculating
the tree-level contributions to all orders in the chiral expansion.
The application of these results to the process
 $\eta \to \pi^0 \gamma \gamma$ is also performed.
 Here $N_c$ is  the number of colours of the QCD
group.  It is at this level of approximation that this model has been
phenomenologically tested. The Lagrangian of the ENJL model is given by
\ba
\rlabel{ENJL1}
{\cal L}_{\rm ENJL} &=& \overline{q} \left\{i\gamma^\mu
\left(\partial_\mu -i v_\mu -i a_\mu \gamma_5\right) -
\left({\cal M} + s - i p \gamma_5 \right) \right\} q
\nonumber\\&&
\hskip-2cm + \frac{8\pi^2 G_S}{N_c\Lambda^2}{\displaystyle \sum_{a,b}}
\left(\overline{q}^a_R
q^b_L\right) \left(\overline{q}^b_L q^a_R\right)
- \frac{8\pi^2 G_V}{N_c\Lambda^2}{\displaystyle \sum_{a,b}} \left[
\left(\overline{q}^a_L \gamma^\mu q^b_L\right)
\left(\overline{q}^b_L \gamma_\mu q^a_L\right) + \left( L \rightarrow
R \right) \right] \,. \nonumber \\
\ea
Here summation over colour degrees of freedom
is understood, $a,b$ are flavour indices and
we have used the following short-hand notations:
$\overline{q}\equiv\left( \overline{u},\overline{d},
\overline{s}\right)$;  $v_\mu$, $a_\mu$, $s$ and
$p$ are external vector, axial-vector, scalar and pseudoscalar
field matrix sources in flavour space;
${\cal M}$ is the current quark-mass matrix.
For values of the input parameters we use the results of Fit 1 in
\rcite{BBR}. These are $G_S=1.216$, $G_V=1.263$ and a cut-off $\Lambda$ in
the proper time regularization of 1.16 GeV. For the current quark-masses
 we use
$m_u=m_d=$ 3.2 MeV and $m_s=$ 83 MeV.
These are the values that give the physical neutral pion and kaon
masses in this model.  Other phenomenological consequences
can be found in \rcite{BBR,ENJLrevs,Physrep} and references therein.

The amplitude for $\gamma(q_1)\gamma(q_2)\to\pi^0(p_1)\pi^0(p_2)$
 can be written in terms of two
amplitudes $A(s,\nu)$ and $B(s,\nu)$.
We use here the conventions of \rcite{BGS}.
\ba
\rlabel{basis}
T(\gamma(q_1)\gamma(q_2)\to\pi^0(p_1)\pi^0(p_2))&=&
\nonumber \\
 && \hskip-4cm  e^2 A(s,\nu)\left[ (q_1\cdot q_2)
(\epsilon_1\cdot\epsilon_2)-(q_1\cdot\epsilon_2)
(q_2\cdot\epsilon_1)\right]\nonumber\\
 && \hskip-4cm + e^2 4 B(s,\nu)\left[ (q_1 \cdot q_2)
(\Delta\cdot\epsilon_1)(\Delta\cdot\epsilon_2)
+(\Delta\cdot q_1)(\Delta\cdot q_2)(\epsilon_1\cdot\epsilon_2)
\right. \nonumber \\
 && \left. \hskip-4cm - (\Delta\cdot q_2) (q_1\cdot\epsilon_2)
(\Delta\cdot\epsilon_1) - (\Delta \cdot q_1) (q_2\cdot\epsilon_1)
(\Delta\cdot\epsilon_2) \right]\,
\ea
where
\be
s = \left(q_1+q_2\right)^2\, ;\quad
t=\left(q_1-p_1\right)^2\, ;\quad
u=\left(q_1-p_2\right)^2\, ; \quad
\nu \equiv  t-u ; \quad  \Delta=p_1-p_2
\ee
and $\epsilon_{1,2}(q_{1,2})$ are the polarization vectors of
photons 1 and 2. For $p_1^2= p_2^2 $ we have
$
-2 \Delta\cdot q_1 = 2 \Delta\cdot q_2 = t-u $.
The above amplitude is manifestly gauge invariant.
The cross-section in terms of $A(s,\nu)$ and
$B(s,\nu)$ can be found in a simple form in \rcite{BGS},
Section 2.

The calculation of the tree-level contributions at leading
order in $1/N_c$  to $A(s,\nu)$ and
$B(s,\nu)$ in the ENJL model to all orders in the momentum expansion
and quark masses is the
main purpose of this paper. The method used is the same one used in
\rcite{Physrep,BP1} to calculate several three-point functions. Here,
we need $SVV$, $SPP$, and $VVP$ one-loop three-point functions, and $PPVV$
and $PVPV$ one-loop four-point functions and all possible full
two-point functions.  We refer to \rcite{Physrep,BP1} for notation and
a detailed description of the method used. Essentially we calculate
numerically the Green's function
\ba
\rlabel{method}
\Pi_{\mu\nu}(q_2,p_1,p_2) &=& \nonumber \\
&& \hskip-3cm i^3 {\dis \int}
{\rm d}^4 x {\dis \int} {\rm d}^4 y  {\dis \int}
{\rm d}^4 z \, e^{i(-q_2\cdot x + p_1\cdot y+ p_2\cdot z)}
 \langle 0| T \left( P(0)P(x)V_\mu(y)V_\nu(z)\right)|0\rangle
\ea
to leading order in $1/N_c$.
Here  $V_\mu (x)= {\dis \sum _a} Q_a \,
(\bar{q}^a\gamma_\mu q^a)(x)$ is the electromagnetic quark current,
$Q_a$ is the flavour $a$ quark electric charge in units of $|e|$
and $P(x)= i(\bar{q}^b \gamma_5 q^b)(x)$. Then we form the correct
flavour ($b$ in $P(x)$) combinations to obtain the pseudoscalar current that
couples to the  neutral pion.
We calculate all the form-factors in $\Pi_{\mu \nu}(q_2,p_1,p_2)$,
 see Appendix A in \rcite{BGS} for their definition,
and check that they satisfy
Bose relations and gauge invariance in our numerical results explicitly.
We then reduce the pion legs by going on-shell following the procedure
described  in \rcite{BP1}. The photon momenta are also
taken at the mass-shell.

There are two main types of contributions:
The first one is one-loop four-point function
with a constituent quark in the loop connected
to the outside legs via a chain of bubbles. With a bubble we mean a one-loop
constituent quark two-point function, bubbles are then
joined by ENJL four-quark vertices to built chains of bubbles.
We refer to these contributions globally as just the
 four-point contribution. In a bosonized language this would be four-meson
vertices (the one-loop four-point function) coupled to the external sources
($V_\mu(x),P(y)$) by propagators (the chains of bubbles).
The other contribution is, in bosonized language, diagrams with two three-meson
vertices connected by a propagator and the remaining free legs connected
to the external sources  by propagators as before.
These we refer to as three-point contributions and we label them by the
spin-parity of the ``meson''  connecting the two vertices.
For $\gamma \gamma \to \pi^0 \pi^0$  these are states
with either $I^{G}(J^{PC})=0^+ (0^{++})$,
 $0^+ (2^{++})$, and $J^{PC}= 1^{--}, 1^{+-}$, while for
$\eta \to \pi^0 \gamma \gamma$  these are states with either
$I^{G}(J^{PC})=1^- (0^{++})$,
 $0^+ (2^{++})$, and $J^{PC}= 1^{--}, 1^{+-}$. In the ENJL model we
are using, only  $J^P= 0^+$ or $1^-$ structures are present, other structures
could be introduced, for instance, adding operators with extra derivatives
 in \rref{ENJL1} like in \rcite{PP} and/or Dirac structures. Consistently
we use the value of the parameters obtained from a global fit to low-energy
data within this model and thus we expect a good description with just
these structures.

There are then, three non-vanishing contributions in this model:
The four-point one, the three-point scalar one and the
three-point vector one. The vector three-point
contribution is gauge-invariant by itself. The scalar-three point and the
four-point contributions need to be added in order to be chiral and gauge
invariant. E.g. at $s = \nu =0$ and for zero quark masses
the amplitude $A(s,\nu)$ should vanish.
This is equivalent to say that the tree-level contributions in the chiral
limit starts at order $p^6$ for the neutral pion process.
Our numerical result satisfies this which therefore
 provides a non-trivial numerical check on the calculation.
To take out the expected order of magnitude of $A(s,\nu)$ and $B(s,\nu)$,
we define
\be
a(s,\nu) \equiv \left(16\pi^2 f_\pi^2\right)^2 A(s,\nu) \, ,
\qquad b(s,\nu) \equiv \left(16\pi^2 f_\pi^2\right)^2 B(s,\nu)\,.
\ee
{}From an analysis of the
possible terms in the chiral Lagrangian, it follows that
\be
\rlabel{aidef}
a^{(6)}(s,\nu) =  a_1 m_\pi^2 + a_2 s \, ,
\qquad b^{(6)}(s,\nu) =  b_1 \,.
\ee
Deviations from this behaviour are an indication of the size of
the corrections of counterterms beyond order $p^6$ within the ENJL model.
The couplings $a_1$, $a_2$ and $b_1$ are order $N_c^2$ in the large
$N_c$ counting.
\begin{figure}
\begin{center}
\leavevmode\epsfxsize=10cm\epsfbox{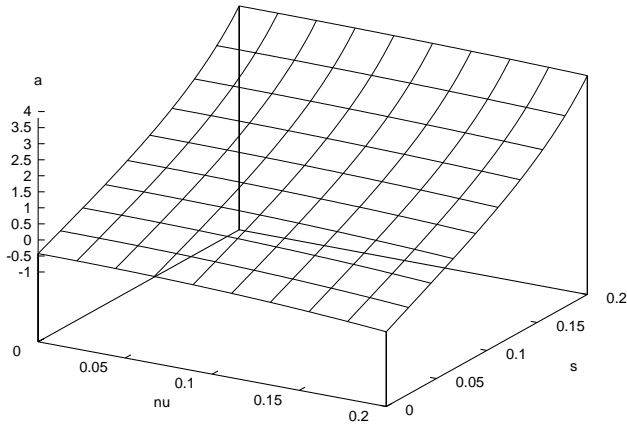}
\end{center}
\caption{The form-factor $a^r(s,\nu)$
in the ENJL model with the experimental
value for the pion mass. The axis $a^r(s,\nu)$, $s$ and $\nu$ are given
in GeV$^2$.}
\rlabel{fig1}
\end{figure}
\begin{figure}
\begin{center}
\leavevmode\epsfxsize=10cm\epsfbox{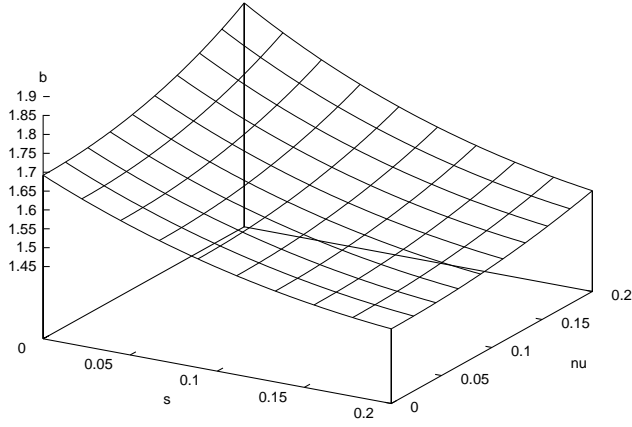}
\end{center}
\caption{The dimensionless
form-factor $b^r(s,\nu)$ in the ENJL model with realistic quark
masses. The axis $s$ and $\nu$ are given in GeV$^2$.}
\rlabel{fig2}
\end{figure}
The amplitude $a^r(s,\nu)$ for the case of real pion mass is shown in
 Fig. \tref{fig1} and $b^r(s,\nu)$ in Fig. \tref{fig2}.
Where the superscript $r$ means the corresponding finite regularized
part. The Bose
symmetry requires them to be symmetric exchanging $\nu$ by $-\nu$.
In both cases $s$ and $\nu$ vary between $0$ and $0.2$ GeV$^2$.
We have only plotted for
this range of $s$ and $\nu$ since we have to stay away from the
two constituent
quark threshold where the artifacts of the ENJL model start dominating
the results.

A good fit to the non-zero pion mass data displayed is given by:
\ba
a^r(s,\nu)&=& - 0.03035 + 0.3773 \, s - 0.8218 \, s^2 - 0.07967\,{{\nu }^2}
     + {{1.626\,s}\over {0.3140 - s}}
\nonumber\\ & +&
0.4307\,  \Biggl[{ {3\, s + 2 \, \nu -0.1458}
\over { 0.3182 + 0.5( s+ \nu )}} +
{{3\, s - 2 \, \nu -0.1458} \over { 0.3182 + 0.5( s - \nu )}}  \Biggr] \, ,
\nonumber\\
b^r(s,\nu)&=& 0.2723 + 0.1604 \, s + 1.798 \, s^2  + 0.0983 \, \nu^2
\nonumber\\&+&
 0.2258 \, \Biggl[{ 1 \over { 0.3182 + 0.5( s+ \nu )}} +
{1 \over { 0.3182 + 0.5( s - \nu )}}  \Biggr] \, .
\nonumber\\
\ea
 For the case with zero quark masses a similarly good fit is:
\ba
\rlabel{fitchi}
a^r_{\chi}(s,\nu)&=& 0.0018 + 0.0919 \, s + 2.380 \, s^2 + 0.13512\,{{\nu }^2}
     + {{1.513\,s}\over {0.2810 - s}}
\nonumber\\ & +&
0.47527\,  \Biggl[{ {3\, s + 2 \, \nu} \over { 0.3364 + 0.5( s+ \nu )}} +
{{3\, s - 2 \, \nu } \over { 0.3364 + 0.5( s - \nu )}}  \Biggr] \, ,
\nonumber\\
b^r_\chi(s,\nu)&=& 0.08432 + 0.3323 \, s + 2.059 \, s^2  + 0.0012 \, \nu^2
\nonumber\\&+&
 0.26426 \, \Biggl[{ 1 \over { 0.3364 + 0.5( s+ \nu )}} +
{1 \over { 0.3364 + 0.5( s - \nu )}}  \Biggr] \, .
\ea
The constraint $a^r_\chi(0,0)=0$ is satisfied by our numerics to about
$10^{-4}$. That \rref{fitchi} deviates by a little more is due to the
quality of the fit. A good fit to the vector contribution alone in
the case of non-zero pion mass is
\ba
a^r_V(s,\nu) &=&  0.08149 - 1.671 \, s + 0.7348 \, s^2 + 0.5022 \,{{\nu }^2}
\nonumber \\ & +&
0.3333\,  \Biggl[{ {3\, s + 2 \, \nu -0.1458}
\over { 0.3182 + 0.5( s+ \nu )}} +
{{3\, s - 2 \, \nu -0.1458} \over { 0.3182 + 0.5( s - \nu )}}  \Biggr] \, ,
\nonumber \\
b^r_V(s,\nu) &=&  - 0.3545 + 0.2566 \, s - 0.1292 \, s^2  - 0.1371 \, \nu^2
\nonumber\\ &+&
 0.1795 \, \Biggl[{ 1 \over { 0.3182 + 0.5( s+ \nu )}} +
{1 \over { 0.3182 + 0.5( s - \nu )}}  \Biggr] \, .
\ea
In all the fits above the form-factor $a^r(s,\nu)$ is in
GeV$^2$ and $b^r(s,\nu)$ is dimensionless.
In these results we have used consistently the large $N_c$
ENJL values for $f_\pi$, i.e.
in the chiral limit $f_\pi= 88.9$ MeV and for the non-zero quark-masses
$f_\pi=$ 90.0 MeV.
It should be remarked that we have chosen a type of meson dominance form
to do the fitting but the values of the poles have no physical meaning.
The expressions just provide a good fit within the kinematical regime
mentioned. The scalar three-point contribution only contributes
to $A(s,\nu)$, not to $B(s,\nu)$. The typical size of the vector
contribution is about half of the total size for both $A(s,\nu)$
and $B(\nu)$ for small $s$ and $\nu$.
 {}From the fits in the chiral limit we can extract $a_2^r$ and $b_1^r$,
  {}from the finite quark mass result we extract $a_1^r$.
The results are in Table \tref{table1}.
\begin{table}[htb]
\begin{center}
\begin{tabular}{|c|c|c|c|c|c|}
\hline
  &   &Resonance&Resonance&Vector& \\
  & ENJL&Exchange &Exchange&Contribution&ENJL\\
  &(this work) &ENJL\protect{\rcite{BB}}&Experiment&ENJL
  &\protect{\rcite{Belkov}}\\
\hline
$a^r_1$ & $-$23.3& $-$20.2 & $-$37.5 $\pm$ 4.5 $\pm$ 4.1 & $-$12.3 & $-$12.1\\
$a^r_2$ & 14.0 & 12.3 &14.4 $\pm$ 2.7 $\pm$ 1.0 & 4.4 & 10.3\\
$b^r_1$ & 1.66 & 1.30 &3.1 $\pm$ 0.24 $\pm$ 0.5 & 0.73&0.97 \\
\hline\end{tabular}
\end{center}
\caption{The comparison of the order $p^6$ part of our result with
meson dominance and  existing ENJL model calculations.}
\rlabel{table1}
\end{table}
The second column is our ENJL result. The third column is the resonance
exchange dominance prediction in Ref. \rcite{BB} within the ENJL
model. In this reference, this result is added to the constituent
quark-loop four-point function contribution.  We believe this is
inconsistent and this procedure is actually adding contributions from
two different models: resonance exchange dominance and the quark-loop
model.  In fact, the four-point function alone does not fulfill chiral
symmetry as said before while the resonance exchange contribution does
by construction.  The comparison of the second and the third column shows
that the resonance exchange dominance works in the ENJL to order $p^6$
within 15$\sim$25 \% similarly to what happened to other quantities at
order $p^4$\rcite{BBR}.  The four-point and the three-point
contributions combine to do this rather well.  This is quite
important, since contrary to the order $p^4$ \rcite{EGPR} this is not
well established yet and our result can be used as support for the use
of resonance exchange dominance to this order as well. In the fourth
column we show the result of resonance exchange dominance using
experimental inputs. They include, of course, higher than order $p^6$
corrections due for instance to quark masses and next to leading in
$1/N_c$ corrections. Here we have
consistently used the experimental value $f_\pi=$ 92.4 MeV.  For the
contribution of the states with $I^G(J^{PC})=0^+(0^{++})$ we have
taken the signs favoured by phenomenology, see \rcite{Jetter}, and
predicted also by the ENJL model \rcite{BB}.  The first error shown
is the one from the input values and the second is the contribution
from the states whose sign is not well established; i.e.
$I^G(J^{PC})=0^+(2^{++})$ \rcite{BGS}.  As can be seen from the table
only the ENJL result for $a^r_2$ is compatible with the corresponding
resonance exchange result. For the other two, although they only
differ by one or two $\sigma$s, the central values are not quite
compatible. The discrepancy is mainly because the decay rates for
$\rho, \omega \to\pi^0\gamma$ are not well reproduced by the ENJL
model\rcite{Ximo}.  The ENJL model does reproduce
$\rho^+\to\pi^+\gamma$ decay rate well\rcite{Ximo}. At present, these
reported radiative vector meson decays do not agree with nonet
symmetry and therefore this discrepancy is not solvable within our
approach.  In the fifth column we show the contribution of the vector
three-point function type of contributions to the results in the
second column.  In the sixth column we show the results obtained in
Ref. \rcite{Belkov}.  We disagree with
\rcite{Belkov} but a look at the table makes it clear that in that
reference the contribution from the vector part (our fifth column) was
neglected.  In view of the importance of this contribution, the
approximation used there is not valid.
\begin{figure}
\begin{center}
\leavevmode\epsfxsize=10cm\epsfbox{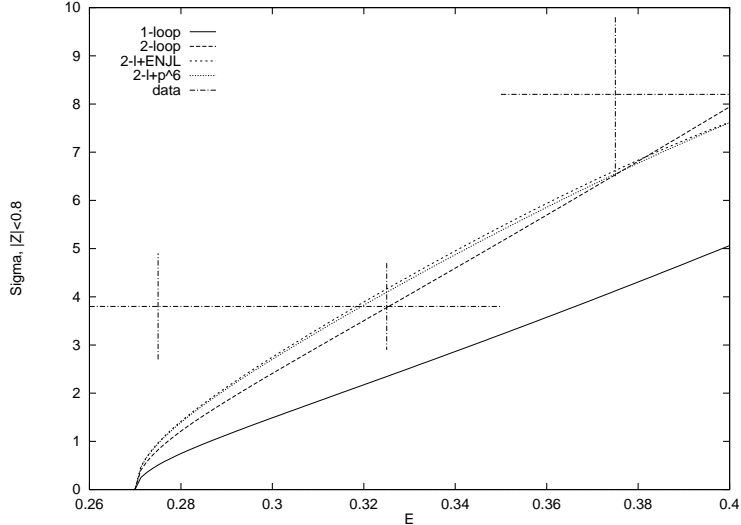}
\end{center}
\caption{The cross-section for $\gamma\gamma\to\pi^0\pi^0$ with $|\cos\theta|
\le 0.8$. Plotted are the result to order $p^4$;
the result to order $p^6$ adding only the two-loop result,
 the result to the same order adding also
the full ENJL result or  adding  only the order $p^6$ part.}
\rlabel{fig3}
\end{figure}

In Fig. \tref{fig3} we have plotted
for the cross-section for $\gamma \gamma \to \pi^0 \pi^0$
the one-loop result,
the two-loop result
with all the order $p^6$ counterterms set to zero and the two-loop result
with the full ENJL contribution added.
 The difference between the last two
curves show the effect of the counterterms. As a comparison we have also
plotted
the result with only the order $p^6$ part of the ENJL result plus the
two-loop result.
For the two-loop result we have used the simplified formula as given in
\rcite{BGS} with $\Delta_A=\Delta_B=0$.
We have also indicated the presently available data\rcite{CB}
in this figure. The integration over the azimuthal angle  was done up to
$|\cos\theta| \le 0.8$ in this figure.
{}From the formulas given above and the expressions in \rcite{BGS}, the
extension to the full integration range can be done easily. As can also be
seen the total effect of the order $p^6$  is quite small
and the effect of the order $p^8$ and higher orders is extremely small. So
up to the energies shown, only a crude estimate of the extra
counterterms is sufficient for this process.

The other decay $\eta\to\pi^0\gamma\gamma$ is more difficult to treat to all
orders in momenta. The problem is that our whole approach is leading
in $1/N_c$. The pseudoscalar mass eigenstates there, do not
correspond to the physical $\eta$ and $\eta^\prime$ since the
U(1)$_A$-breaking due to the fact that the
anomaly is not present at this order. Therefore
we cannot directly calculate the relevant $\eta$ amplitude as we did
for $\gamma \gamma \to \pi^0 \pi^0$. We could
 resort to calculate the $\eta$ decay at tree-level to all orders in CHPT
 leading in $1/N_c$
if we could obtain the relevant  counterterms allowed by the symmetry
 {}from our all orders calculation of $\gamma \gamma \to \pi^0 \pi^0$.
The physical $\eta$ field in terms of the nonet $\eta_8$ and singlet
$\eta_1$ SU(3) flavour states is $\eta = \cos\varphi_P \, \eta_8-
\sin\varphi_P \, \eta_1$. We have used $\sin\varphi_P = -1/3$ \rcite{PDG94}.
Using the same basis as in Eq. \rref{basis} but substituting
 $p_1=p_\eta$ and $p_2=-p_\pi$ everywhere, the amplitude $\eta \to
 \pi^0 \gamma\gamma$ can be written in terms of $A(s,\nu)$ and
 $B(s,\nu)$ which to order $p^6$ and large $N_c$ to be consistent with
 our calculation of the amplitudes $A(s,\nu)$ and $B(s,\nu)$,
can be parametrized as:
\ba
\rlabel{etap6}
A^{(6)}(s,\nu)&=& {{4}\over{3 f_\pi^4}} \, \sqrt{2 \over3} \,
 \left[ 8 m_\pi^2 (2 d_3^r - d_2^r) - 8 m_\eta^2 d_2^r +
  (d_1^r + 8 d_2^r) s \right] \, ; \nonumber \\
B^{(6)}(s,\nu)&=& - {{2}\over{3 f_\pi^4}} \, \sqrt{2 \over3} \, d_1^r \,.
\ea
Here, $d_1^r$, $d_2^r$ and $d_3^r$ are the order $p^6$ couplings
of the effective chiral Lagrangian defined
in  Eq. (11) of \rcite{BB}\footnote{Notice we disagree with the result
for $A^{(6)}(s,\nu)$ in that reference.}. They are order $N_c^2$
in the large $N_c$.
These couplings also coincide with the ones
defined in \rcite{BGS} when restricted to the two flavour case.
In general (next-to leading in $1/N_c$) there are three more couplings
\rcite{Jetter} which are order $N_c$. Only two of them appear in the
amplitudes $A(s,\nu)$  and $B(s,\nu)$ to order $p^6$  and can
be seen as $1/N_c$ corrections  of the $d^r_1$ and $d^r_2$  couplings in
Eq. \rref{etap6} \rcite{Jetter}.
 These same $d^r_{1,2,3}$ couplings  enter  in the
order $p^6$  expression for the amplitudes $A(s,\nu)$ and $B(s,\nu)$
for $\gamma \gamma \to \pi^0 \pi^0$, see
\rcite{BGS,BB}. The fact that they appear there in three different
combinations
allows us to disentangle them completely from  the different fits
to the ENJL data shown before.
To obtain all  counterterms contributing to the $\eta$ decay
{}from $\gamma \gamma \to \pi^0 \pi^0$
at higher orders is not possible.  The underlying problem is easy to
understand. Due
to relations like $ 2 q_1 \cdot q_2  = 2 p_1 \cdot p_2 + p_1^2
+ p_2^2$,  the combinations of counterterms appearing in the amplitudes
 for the $\eta$ decay and   $\gamma \gamma \to \pi^0 \pi^0$
are clearly different. Therefore
the only possibility to make a prediction to all orders for the
 $\eta$ decay is to determine all the
couplings modulating the needed counterterms.
However,  a quick counting shows that the  number of different combinations of
counterterms appearing at higher order ($p^8$
and higher) does not allow to determine all possible terms in
the chiral Lagrangian. It is enough however, as said before, for
the order $p^6$ counterterms.
One could hope that going to the off-shell parts of the four-point
function in
\rref{method} it could be done, but this is not the case.
 There are  again terms
that contribute to $\gamma \gamma \to \pi^0 \pi^0$ off-shell differently as
to the $\eta$ decay
and terms that contribute to \rref{method}
but not to the decays. An example of the latter is
\be
\mbox{tr}\left(F_{\mu\nu}F^{\mu\nu}
\chi\chi^\dagger\right)\qquad\mbox{with}\qquad \chi=2B_0(s+ip)\,.
\ee
where $s$ and $p$ are  scalar and  pseudoscalar external sources
 as in \rref{ENJL1}. And, of course, the number of different
combinations at order $p^8$ and higher is not enough to disentangle
all possible terms in the chiral Lagrangian.
For this reason we will only use the order $p^6$ part of the
calculation for the $\eta$-decay.

{}From Refs. \rcite{BGS,BB} one can obtain the relation between
the couplings
$a_1$, $a_2$ and $b_1$ in \rref{aidef} and the $d_{1,2,3}$
couplings of the order $p^6$ Lagrangian,
\ba
\rlabel{rela}
d_1 &=& -  {9 \over 10} {1\over {(16 \pi^2)^2}} \, b_1 \\
d_2 &=&   {9 \over 160} {1\over {(16 \pi^2)^2}} \,
\left[a_2+2 b_1 \right] \\
d_3 &=&   {9 \over 320} {1\over {(16 \pi^2)^2}} \,
\left[a_1 +2 a_2+4 b_1 \right] \, .
\ea
{} From our ENJL results in the second column of
Table \rref{table1} we get
\be
\rlabel{enjld}
d^r_1 = -6.0 \cdot 10^{-5} ; \qquad
d^r_2 = 3.9 \cdot 10^{-5} ; \qquad
d^r_3 = 1.3 \cdot 10^{-5} .
\ee
We observe that the three couplings are of the same order of
 magnitude in this ENJL model. Notice that at order $p^6$
there is nonet symmetry and the difference between isospin
one and isospin zero is higher order.

One can also use resonance exchange dominance in the
eta decay to predict the $d^r_i$ couplings. This has been done before
in \rcite{ABBC} for the $d^r_{1,2}$ and in \rcite{Ko}
to predict also $d^r_3$ using $a_0(980)$ data. Since the actual data on
the $a_0(980)$ \rcite{PDG94} do not allow to make any trustable
prediction,
 we have used nonet symmetry to obtain the $d^r_3$ coupling from
the fourth column in Table \tref{table1} with the formulas
in \rref{rela}. We get
\be
\rlabel{resonanced}
d^r_1 = - (8.2 \pm 2.0)\cdot 10^{-5} ; \qquad
d^r_2 = (4.3  \pm 1.0)\cdot 10^{-5} ; \qquad
d^r_3 = (0.4 \pm 2.7) \cdot 10^{-5} .
\ee
The $d^r_i$ couplings obtained in
\rref{resonanced} contain higher than order $p^6$ corrections
  due to quark masses contributions to the masses of the resonances
as well as next-to-leading in $1/N_c$ corrections.
 Keeping this in mind, we observe that only $d^r_2$ is in complete
agreement although
both results are  compatible at the one $\sigma$ level.
Notice the large error bars in this way of estimating
the $d^r_i$ couplings.

The decay rate  for $\eta \to \pi^0 \gamma \gamma$
can be written as follows
\ba
\Gamma (\eta \to \pi^0 \gamma \gamma)
&=& {\alpha^2   \over {64 \pi m_\eta^3}}
{\dis \int^{s_2}_0} {\rm d} s \, s^2 \,  {\dis \int^{t_2}_{t_1}} {\rm d} t
\, \left[   \left| H_{++}(s,\nu) \right|^2 +\left|  H_{+-}
(s,\nu) \right|^2 \right] \,
\ea
with $\alpha$ the fine structure constant and
\ba
s_2&=&(m_\eta-m_\pi)^2,  \nonumber \\
t_{2,1} &=& {1\over 2} \left[ m_\eta^2+m_\pi^2-s \pm
\sqrt{(m_\eta^2+m_\pi^2-s)^2-4m_\eta^2 m_\pi^2} \right]
\ea
and
\ba
H_{++}(s,\nu)&=& A(s,\nu) + 2 B(s,\nu) \, ( 2 m_\pi^2
+ 2 m_\eta^2 - s) \, ; \nonumber \\
H_{+-}(s,\nu)&=& 8 B(s,\nu) \, {{m_\pi^2 m_\eta^2 - u t}
\over {s}} \, .
\ea
The experimental value of the decay rate  $\Gamma(\eta \to
\pi^0 \gamma \gamma)$ is
 \be
\Gamma(\eta \to \pi^0 \gamma \gamma) =
(0.85 \pm 0.19) \, {\rm eV} \protect{\rcite{PDG94}}.
\ee
Using only the order $p^6$ tree-level contributions
to $A(s,\nu)$ and $B(s,\nu)$ in \rref{etap6} we get
\be
\Gamma(\eta \to \pi^0 \gamma \gamma) = 0.18 \, {\rm eV}
\ee
using our ENJL results for the $d_i^r$ couplings in \rref{enjld} and the
chiral limit ENJL value for $f_\pi$, i.e. $f_\pi=$ 88.9~MeV.
The contribution of the coupling $d_3^r$ is not dominant and its
ENJL value results in a decreasing of the the decay rate with respect
to the case with
$d_3^r=0$. If one instead uses  the values in \rref{resonanced}
then
 \be
\Gamma(\eta \to \pi^0 \gamma \gamma) =
(0.18 ^{+0.15}_{-0.10}) \, {\rm eV}.
\ee
Notice that the variation within the allowed range of values
for $d^r_{1,2,3}$ predicted by the resonance exchange model
produces a large uncertainty in the decay rate.
 This uncertainty is avoided in the ENJL model predictions.

The contribution from the order $p^4$ loops is either suppressed
by G-parity or the kaon mass \rcite{ABBC}. The analysis in \rcite{Jetter}
of the  order  $p^6$ loop contributions, though partial, shows the
same suppression. The order $p^6$ contributions included there,
which are expected to be the dominant ones at that order,
interfere destructively decreasing the decay rate but only by 0.04 eV.
At order $p^8$ there appears  qualitatively new contributions
\rcite{ABBC}, they are the doubly-anomalous contributions. Its
relative size compared to the chiral loop contributions analysed previously,
 cannot be inferred from the
 chiral counting since it is the first in its class of contributions.
There is, for instance, no G-parity suppression in the couplings
\rcite{ABBC}.

So,  adding the order $p^4$ charged pion and kaon loop contributions
plus the doubly-anomalous contributions of order $p^8$ and the
tree-level order $p^6$ contributions
to $A(s,\nu)$ and $B(s,\nu)$, we get
\be
\Gamma(\eta \to \pi^0 \gamma \gamma) = 0.30 \, {\rm eV}
\ee
using our large $N_c$
ENJL results for the $d_i^r$ couplings in \rref{enjld}.
If we instead use the values in \rref{resonanced}
one obtains
 \be
\Gamma(\eta \to \pi^0 \gamma \gamma) =
 (0.27 ^{+0.18}_{-0.07}) \, {\rm eV}.
\ee
There is a very strong constructive interference between the
order $p^6$ tree-level and the order $p^4$ and $p^8$ loop
contributions. Including the expected small decreasing
of the order $p^6$ loops, we conclude that within the
ENJL model, the order $p^6$ prediction for the decay rate $\Gamma(\eta
\to \pi^0 \gamma \gamma)$ is off the experimental result
by almost three  $\sigma$s. This is in disagreement with the
results in \rcite{BB}.

In the present work, we have  computed the tree-level contributions
to all orders in the chiral expansion and leading
in $1/N_c$ for $\gamma \gamma \to  \pi^0 \pi^0$ within the ENJL model.
 We have predicted the corresponding cross-section and compared with
experiment. Our result shows that tree-level contributions of order
 higher than $p^6$ are negligible for $s$ and $\nu$ below 0.2 GeV$^2$.
 We have also predicted the order
$p^6$ counterterms that contribute at large $N_c$ to this process.
A comparison with other   estimates of these counterterms is  made.
We have seen that the resonance exchange dominance works within
the ENJL model  to 15$\sim$25\%. For the $\eta
\to \pi^0 \gamma \gamma$ we have made a prediction  including
the dominant chiral loop corrections \rcite{ABBC,Jetter}, i.e
 those of order
$p^4$ and the doubly-anomalous of order $p^8$, and the tree-level
order $p^6$ obtained within the ENJL model at leading order in $1/N_c$.
We obtain a three $\sigma$s discrepancy with the experimental result
previously observed in other estimates.
We do not expect to obtain
unusually large corrections for $A$ and $B$ from higher order terms
because of the CHPT counting. This is  the case,
for instance, for the order $p^6$ loops analysed in \rcite{Jetter}.
It should be noticed, however, that a small change in the values
of $A$ and $B$
can result in a large enhancement of the decay rate.
An example is the enhancement due to the
strong constructive interference between the leading loop correction
and the tree-level contributions.
This is also well illustrated by the increasing of 0.14 eV
(almost one $\sigma$)
when higher order tree-level contributions are taken into
account  by an ``all-orders'' vector meson resonance  exchange model
\rcite{ABBC}. This would bring our order $p^6$ ENJL estimate
closer to the experimental result, but still off the experimental
result by not less than two $\sigma$s.  As said in the text,
the next to leading in $1/N_c$ couplings could be regarded
as corrections to $d^r_1$ and $d^r_2$ \rcite{Jetter}. These
have to be added to the errors inherent in our large $N_c$ ENJL
model. In view of the large constructive
interference mentioned above they could add a significant contribution
to the decay rate. In fact, reasonable $1/N_c$ corrections (30~\%)
 together with the higher orders effect above
could easily bring the final result within one $\sigma$ from the
experimental result.
It is then of interest to have a high statistics measurement
of this decay rate and the two-photon energy spectrum in order
to reduce the actual experimental uncertainty and
see if this discrepancy persists. It could be also used to extract
the effective $d^r_i$ couplings and therefore  deviations  from our
large $N_c$ estimate. Due to its large present uncertainty,
the decay rate $\Gamma(\eta \to \pi^0 \gamma \gamma)$
can within  one standard deviation be explained with higher order corrections
both in $1/N_c$ and CHPT.

\section*{Acknowledgements}
This work was partially supported by NorFA grant 93.15.078/00.
JB and JP  are grateful to
the Benasque Center for Physics where part of this work was done.
JP is indebted to the Niels Bohr Institute and NORDITA where
most of his work was done for support, he also thanks
CICYT (Spain) for partial support under Grant Nr. AEN93-0234
and the DESY Theory group where part of his work was done for
hospitality.

\end{document}